\documentstyle[preprint,aps]{revtex}
\begin{document}
\draft
\title{Character of electron reflection at a normal metal-Peierls
semiconductor boundary}
\author{S. N. Artemenko and S. V. Remizov}
\address{
Institute for Radioengineering and Electronics of the Russian Academy of
Sciences, 103907 Moscow, Russia}
\date{\today}
\maketitle

JETP Letters, Vol. 65, No. 1, 53-58 (10 January 1997)

\abstract
The reflection of electrons incident from a normal metal on the boundary of
the metal with a quasi-one-dimensional conductor containing a charge-density
wave (CDW) is investigated theoretically. It is shown that the reflection is
not of an Andreev character but rather of a Bragg character. This is due to
the fact that the CDW is actually an electronic crystal, and its wave vector
is a reciprocal lattice vector of the electronic crystal. The ratio of the
intensities of the standard and Bragg reflection depends on the phase of the
CDW.
\endabstract

It is well known (see, for example, Ref. 1) that an electronic crystal -- a
charge or spin density wave whose motion under the influence of an electric
field is associated with a collective mechanism of conduction -- forms in
quasi-one-dimensional conductors below the Peierls transition temperature.
For definiteness, in what follows we shall study a charge-density wave (CDW),
but the results obtained are also applicable to the case of a spin density
wave.

There exists a formal analogy between Peierls semiconductors (PSs) and
superconductors, since in both cases the condensed state is described by an
order parameter $\Delta = |\Delta|\exp{i\varphi}$ whose amplitude determines
the energy gap in the single-particle excitation spectrum and the derivative
of whose phase (in a superconductor with respect to the coordinate and in a
Peierls semiconductor with respect to time) is proportional to the
contribution of the condensed electrons to the electric current density. A
CDW can be graphically imagined as a condensate consisting of bound pairs of
electrons and holes whose momenta differ by the magnitude of the wave vector
of the CDW. By analogy to superconductors, where the condensate consists of
pairs of electrons with opposite momenta and Andreev reflection is observed
at a boundary with normal metal \cite{An}, it should be expected that even the
reflection of electrons with energy close to the Fermi energy from the normal
metal-PS boundary $(N-P)$ has an unusual character. It has been concluded in
theoretical works \cite{KP1,KP2} that an electron relected from a PS onto
which it was incident from a normal metal moves along the same trajectory
along which it was incident on the PS, i.e., the reflection is similar to
Andreev reflection, but in contrast to a superconductor the sign of the
charge of the incident quasiparticle does not change. The observation of
features in the resistance of a contact of a PS with a normal metal which
were interpreted as a manifestation of the Andreev-type reflection predicted
in Refs. 3 and 4 was recently reported in Ref. 5. In our view,
a reflection in which the reflected particle moves along the same trajectory
cannot appear at a metal-PS contact, since a quasiparticle in the PS is a
superposition of two electrons with momenta differing by the wave vector of
the CDW and not with opposite momenta. We shall show that the momentum
component parallel to the interface can either be conserved (standard
reflection) or change by an amount equal to the CDW wave vector component
parallel to the interface (Bragg reflection from an electronic crystal).

We shall be interested in the reflection of electrons with energies of the
order of $k_B T$ or $\Delta$ near the Fermi energy, since such electrons
will determine the conductivity in structures containing a PS.

We consider first the reflection of electrons at the interface between a
normal metal and a PS, whose electronic structures differ from one another
only by the presence of a CDW in the PS, occupying the space $x > 0$. This
model will enable us to investigate reflection from a CDW in a pure form,
since there will be no reflection, associated with the difference in the
energy structure of the crystals and having no relation with the CDW, from
the interface. The fact that the CDW was formed in only a part of the crystal
could be due to the fact, for example, that the electron-phonon interaction
constant vanishes for $x < 0$.

To calculate the electronic wave functions in a PS it is often convenient to
employ the self-consistent field approximation equations of the Bogolyubov-de
Gennes type for superconductors, as was done, for example, in Ref. 3. The
envelopes $u({\bf r})$ and $v({\bf r})$ -- the amplitudes determining the
contribution of the states belonging to opposite sheets of the Fermi level,
shifted by the wave vector ${\bf Q}$ of the CDW, to the total wave functions
-- serve as the elements of the spinor wave functions.  In calculating the
wave functions in a nonuniform system by matching at the interface, the total
wave functions
\begin{equation}
\psi =u e^{i{\bf Qr}/2} + v e^{-i{\bf Qr}/2}, \label{ps}
\end{equation}
which are solutions of the Schr\"{o}dinger equation with the potential of the
CDW, prescribed for $x > 0$ as $2|\Delta|\cos{({\bf Qr}+ \varphi )}$, must be
matched. The matching of the envelopes $u$ and $v$ at the interface in the
case of a CDW gives, generally speaking, an incorrect result.

Let the conducting chains be directed along the $x$ axis and let the
electronic spectrum of the quasi-one-dimensional conductor in the normal
state have the form $E_N = p_x^2/2m + E_\perp ({\bf p}_\perp)$, where
$|E_\perp| \ll E_F$ and $E_F = p_{xF}^2/2m$ is the Fermi energy. For
definiteness, we shall consider the case when the wave vector of the CDW
possesses the components ${\bf Q}=(2k_F, Q_y, 0)$. Then in the Peierls state
the spectrum has the form $E_P = \eta \pm \sqrt{\xi^2 + |\Delta|^2}$, where
$\xi = [E ({\bf k})- E ({\bf k}-{\bf Q})]/2$, $\eta = [E ({\bf k})+ E({\bf
k}-{\bf Q})]/2$, and the relation between $u$ and $u$ in Eq. (1) is
determined by $u=-\Delta v /(\xi +\sqrt{\xi^2 + |\Delta|^2})$. If the energy
is measured from the Fermi energy $|\epsilon| = |E - E_F|$ is less than
$|\Delta|$, then in the state with the CDW $|u|=|v|$ to within corrections of
order $\Delta/E_F$. It follows from the form of Eq. (1) that as a result of
matching the solutions at $x = 0$ it will turn out that, to the same
accuracy, the amplitudes $u$ and $u$, which in the normal metal correspond to
the amplitudes of the incident and reflected waves, are of the same absolute
magnitude. Therefore if the wave vector of the CDW possesses a component
$Q_y$ parallel to the interface, then the corresponding component of the
electron momentum will change by $Q_y$ on reflection.

We shall now calculate, by matching the wave functions, the reflection
coefficient, neglecting the coordinate dependence of the energy gap near the
interface as a result of the proximity effect. Let us assume, for simplicity,
that on formation of a CDW the period is doubled in a direction perpendicular
to the conducting chains, i.e., $2Q_y$ corresponds to a reciprocal-lattice
vector. For $x<0$ we seek the wave function in the form
\begin{equation}
\psi = \left[e^{ik_x x } + A e^{-ik_x x} + B e^{-ik_x x +iQ_y y} \right]
e^{i(k_y y +k_z z)}, \label{p-}
\end{equation}
where the first term describes the incident wave, the second term describes
the standard reflection, and the third term describes Bragg reflection. For

$x>0$ the wave function must be a linear combination of functions of the form

(\ref{ps}), which describe states possessing along the y axis momentum
components $k_y$ and $k_y + Q_y$ and the same energy as the state (\ref{p-}).
Strictly speaking, the wave functions in the form of plane wave combinations
considered above can be equated only if the electronic structure in the N and
P regions is the same and the Bloch periodic factors are identical for $x>0$
and $x<0$. Nonetheless, to understand qualitatively the effect of the
difference of the electronic spectra in the $N$ and $P$ regions, we shall
also discuss the result of the matching for the case when the electronic
spectra on both sides of the interface are different.

The expressions for the reflection coefficient $R$ in the general case are
quite complicated. For this reason, we confine our attention to the limiting
case of weak threedimensionality of the spectrum in the conductor with the
CDW and we neglect terms of the order of $E_\perp/\epsilon$. For the case
when the electronic structure is the same to the left and right of the
boundary, we obtain for the ratio of the standard and Bragg reflection
intensities
\begin{equation}
|A/B|^2=(|\Delta|\sin{\varphi}/E_F)^2, \label{a1}
\end{equation}
Therefore, in accordance with what we have said above, $|A| \ll |B|$ and
Bragg reflection, where the parallel component of the momentum changes by
$Q_y$ , dominates. We also note that the relation (\ref{a1}) depends on the
phase of the CDW. $|\epsilon| < |\Delta|$ the reflection coefficient $R=1$,
and for $|\epsilon| > |\Delta|$ we obtain $R=|\Delta|^2/(|\epsilon| +
\xi)^2$, where $\xi = \sqrt{\epsilon^2 - |\Delta|^2}$.

We now consider the case when the effective masses along the $x$ axis are
different in the materials to the left and right of the interface. Then

\begin{equation}
|A/B|^2= \frac{(m_1-m_2)^2}{4m_1m_2} \left\{ \begin{array}{cc}
\sin^2{(\varphi+\varphi_0)} & \mbox{при}\;\; |\epsilon| <|\Delta|,\\
(\epsilon/|\Delta|)^2 - \cos^2{\varphi} & \mbox{при}\;\; |\epsilon| >|\Delta|,
\end{array}
\right.  \label{a2}
\end{equation}
where $\varphi_0= \arctan{\xi/\epsilon}$.For $m_1=m_2$ when the answer in Eq.
(\ref{a2}) vanishes, the terms of order 1 in the ratio $|A/B|^2$ cancel and
therefore the small terms of the order of $(|\Delta|/E_F)^2$, which result in
the formula (\ref{a1}), must be taken into account. Therefore, when the
electronic spectrum of the crystals on different sides of the interface is
different, the standard reflection, where the angle of incidence equals the
angle of reflection, is added to the Bragg scattering, and the intensities of
both types of scattering are of the same order of magnitude and their ratio
depends on the phase of the CDW. If it is assumed that an isotropic metal
fills the space $x<0$, then one obtains an expression differing from
expression (\ref{a2}) by the fact that the effective mass $m_1$ in the $N$
region is replaced by $m_1/\cos{\theta}$, where $\theta$ is the angle of
incidence of the electron. Of course, this result is not quantitative, since
in matching the wave functions we neglected the periodic Bloch factors in
them. Taking account of the real crystal structure would have resulted in the
appearance of terms in the expansion of the wave function in a Fourier series
in the coordinate which correspond to a change in the momentum by an
arbitrary reciprocal lattice vector of both the main and electronic crystals,
as a result of which reflection would have also appeared in other directions
corresponding to Bragg scattering.

Since the phase of the CDW can change when an electric field directed along
the conducting chains is applied to the PS, the dependence of the character
of the reflection on the phase of the CDW can be used for experimental
investigation of $N-P$ contacts.

We note one other interesting feature of reflection from an $N-P$ contact.
This feature is reminiscent of the properties of a normal
metal-superconductor contact. As is well known, in a thin layer of normal
metal of thickness $d$ bordering a superconductor, bound states with an
energy splitting of the order of $\varepsilon_0 =\pi \hbar v_F/d$ (here $v_F
= p_{xF}/m)$) appear at energies $|\epsilon| < |\Delta|$ as a result of the
Andreev reflection from the superconductor.  A similar quantization also
appears in a normal metal in contact with a PS if the thickness $d$ of the
normal metal is less than the mean-free path length. The simplest method for
investigating such quantization is to calculate the density of states with
the aid of the quasiclassical equations for the momentum-integrated Green's
functions, which were employed for investigating the transport properties of
PS\cite{A1,A9}. For our purposes it is sufficient to solve an equation for
the retarded Green's function neglecting the collision integral, in which
case this equation has the very simple form
\begin{eqnarray}
i\hbar v\frac{d\hat{g}}{dx}+ (\tilde{\epsilon}\sigma_z + \hat{\Delta})
\hat{g}-\hat{g} (\tilde{\epsilon}\sigma_z + \hat{\Delta}) =0, \label{g}
\end{eqnarray}
where the Green's function $\hat{g}$ is a $2 \times 2$ matrix with respect to
the index corresponding to opposite sheets of the Fermi surface which are
displaced by the wave vector ${\bf Q}$ of the CDW, $\tilde{\epsilon}=
\epsilon - \eta\,({\bf p}_\perp)$, $\hat{\Delta} =i |\Delta| (\sigma_y
\cos{\varphi} + \sigma_x \sin{\varphi})$, and $\sigma_\alpha$ are the Pauli
matrices. Assuming once again that $|\Delta|$ vanishes abruptly for $x<0$
and that the normal metal occupies the region $-d<x<0$, we solve Eq.
(\ref{g}) with the boundary condition ${\rm Tr}\,\sigma_x g(0)=0$ at the
boundary of the normal metal with the vacuum. From Eq. (\ref{g}) we obtain
for the function $g ={\rm Tr}(\sigma_z \hat{g})$, whose real part
determines the density of states, in the normal region
\begin{equation}
g= (\xi + i \tilde{\epsilon} t)/(\tilde{\epsilon}t +i\xi),  \label{gz}
\end{equation}
where $t=\tan{(2\tilde{\epsilon}d/\hbar v_F + \varphi)}$ and it must be
assumed that $\xi$ is an analytic function of $\epsilon$ in the upper
half-plane. The off-diagonal components of $\hat{g}$ do not vanish even in
the normal region, where they oscillate as $\exp{(2\tilde{\epsilon}d/\hbar
v_F)}$, since the decay length for them equals the mean-free path length and
is greater than the thickness of the normal region. In an accurate
calculation, we would have to take account of the lowering of the energy gap,
which changes the shape of the potential well, in the PS at distances of the
order of $\hbar v_F/|\Delta|$, which happens as a result of the proximity
effect. This lowering is due to perturbations of the off-diagonal components
of $\hat{g}$ in the region of the PS near the contact, but the change in the
shape of the potential well does not affect the qualitative conclusions and
we shall neglect it.

One can see from Eq. (\ref{gz}) that the density of states in the normal
metal is an oscillating function of the phase of the CDW as well as of the
energy and thickness of the normal region. For energies $|\epsilon| <
|\Delta|$ at which bound states appear the density of states has the form
\begin{equation}
N(\epsilon)=\pi \langle (|\xi| +  \tilde{\epsilon} t)
\delta\,(\tilde{\epsilon}t -|\xi|) \rangle. \label{N}
\end{equation}
Here $\langle ... \rangle$ indicates averaging over ${\bf p}_\perp$. At
energies much less than $|\Delta|$ formula (\ref{N}) reduces to
$$N(\epsilon)=\varepsilon_0 \langle
\delta\,(\tilde{\epsilon} -(2n+1)\varepsilon_0) \rangle,$$
where $n$ is a positive integer.

Therefore, according to Eq. (\ref{N}), the electron spectrum in the $N$ layer
consists of bands of width $E_\perp$ (the width, determined by the function
$\eta\,({\bf p}_\perp)$, of the electron energy band in the perpendicular
direction). If the electron spectrum is strongly one-dimensional $E_\perp <
\varepsilon_0$, then the allowed energy bands are separated by regions of
forbidden energies, otherwise they overlap and the energy dependence of the
density of states is of a stepped form. An estimate gives $\varepsilon_0
\approx 70$ K for $v_F = 3 \cdot 10^7$ cm/s and $d = 0,1 \mu$m. At sufficiently
low temperatures ($T <\varepsilon_0, \;E_\perp $) energy quantization can be
observed in measurements of the conductivity of the normal region in the
contact plane or in measurements of the density of states performed with the
aid of tunneling or point contacts to the normal region. The application of
an electric field in the direction of the chains changes the phase of the
CDW, and for this reason, according to Eq. (\ref{N}), the field should
influence the density of states. To observe quantization, the width of the
contact should not exceed the phase correlation length, since the phase of
the CDW depends on the coordinates as a result of impurity pinning.

It was assumed above in the analysis of quantization that with the exception
of the presence of the CDW for $x>0$ the electronic spectrum is the same on
both sides of the contact. The effect of the differences in the electronic
spectrum can be estimated by matching the wave functions, as done in the
investigation of reflection. It is found that in the case of a contact of two
different quasi-one-dimensional metals with $|k_{xF}^{N} -k_{xF}^{P}| \ll
|k_{xF}^{N} +k_{xF}^{P}|$, where $k_{xF}^{N}$ is the Fermi wave vector in the
normal region and $k_{xF}^{P}$ is the Fermi wave vector in the PS, the
quantization condition is obtained if the argument of the tangent
$2\tilde{\epsilon}d/(\hbar v_F)$ in formula (\ref{N}) is replaced by
$\tilde{\epsilon}_N d /(\hbar v_F) + 2(k_{xF}^{N} -k_{xF}^{P})d$, where the
dependence of the energy on ${\bf p}_\perp$ in the normal region appears in
$\tilde{\epsilon}_N$ in the same way as in the similar dependence in
$\tilde{\epsilon}$ for a PS.  Therefore the thickness of the $N$ region
enters the quantization condition not only in the form 60 but also in the
form of the product $(k_{xF}^{N} -k_{xF}^{P})d$, which for large differences
in $k_{xF}$ should blur the quantization effects even as a result of small
variations of the thickness $d$.

Thus we have found that when electrons incident from a normal metal onto the
interface with a PS are reflected, the component of their wave vector
parallel to the interface is either conserved or changes by an amount equal
to the projection of the wave vector of the CDW on the contact plane. The
intensity of different types of reflection depends on the phase of the CDW
and therefore can change when an electric field is ; ;Page 6 ; applied.
Specifically, the reflection of electrons from the region with a Peierls gap
under certain conditions can result in quantization of the energy spectrum of
the electrons near the Fermi energy.

We thank V. A. Volkov for a discussion of this work and also 1. G. Gorlova
and A.  A. Sinchenko for a discussion and for familiarizing us with the
experimental data prior to publication. This work is supported by Russian
Fund for Fundamental Research (Grant 95-02-05392) and MTP "Physics of
Solid-State Nanostructures" (Grant 1-018).

\end{document}